\documentclass{pasj00}
\draft
\usepackage{algorithmic}
\usepackage[ruled]{algorithm2e}

\usepackage{graphicx}
\usepackage{subfig}

\begin{document}
\SetRunningHead{M.Kaplan and H. Sayg\i n}{Dynamic Era-Based Time-Symmetric Block Time-Step Algorithm}

\title{A Dynamic Era-Based Time-Symmetric Block Time-Step Algorithm with 
Parallel Implementations}

\author{Murat \textsc{KAPLAN} }
\affil{Akdeniz University, Space Sciences and Technologies, TR-07058 Antalya,Turkey}
\email{muratkaplan@akdeniz.edu.tr}           

\author{Hasan \textsc{SAYGIN}}
\affil{\.Istanbul Ayd\i n University, Be\c syol Mah. In\"on\"u Cad. No:38
Sefak\"oy-K\"u\c c\"uk\c cekmece,\. Istanbul, Turkey}\email{hasansaygin@aydin.edu.tr}

%


\maketitle

\begin{abstract}
The time-symmetric block time--step (TSBTS) algorithm is a newly developed 
efficient scheme for $N$--body integrations. It is constructed on an era-based 
iteration. In this work, we re-designed the TSBTS integration scheme with dynamically 
changing era size. A number of numerical tests were performed to show the importance 
of choosing the size of the era, especially for long time integrations.
Our second aim was to show that the TSBTS scheme is as suitable as previously known 
schemes for developing parallel $N$--body codes. In this work, we relied on a parallel 
scheme using the copy algorithm for the time-symmetric scheme. We implemented a hybrid of 
data and task parallelization for force calculation to handle load balancing 
problems that can appear in practice. Using the Plummer model initial conditions for 
different numbers of particles, we obtained the expected efficiency and speedup 
for a small number of particles. Although parallelization of the direct $N$--body 
codes is negatively affected by the communication/calculation ratios, we obtained 
good load balance results. Moreover, we were able to conserve the advantages of the 
algorithm (e.g., energy conservation for long--term simulations).
\end{abstract}

\section{Introduction}
\label{intro}
In many practical applications in $N$--body integrations, the block time--step approach
is preferred. In this approach, many particles share the same step size, where 
the only allowed values for the time--step length are powers of two. Block time--steps 
are advantageous to reduce the prediction overheads, and are needed both 
for good parallelization and code efficiency. However, the time-symmetricity and 
symplecticity of previous direct integration schemes are disturbed by using 
variable block time--steps.

The algorithm developed by \citet{Makino-2006} (TSBTS) is the first algorithm 
for time symmetrizing block time--steps which carry the benefits of time 
symmetry to block time--step algorithms. In this algorithmic approach,
the total history of the simulation is divided into a number of smaller 
periods, with each of these smaller periods called an ``era''. Symmetrization is 
achieved by applying a time symmetrization procedure with an era-based iteration.

The TSBTS algorithm was generated for direct integration of $N$--body systems and 
as such is suitable to use for a moderate number of bodies no more than $10^5$. 
The direct approach to $N$--body integration is preferred when we are interested 
in the close-range dynamics of the particles, and aiming at obtaining high accuracy. 
The algorithm gives us the ability to reach long integration times with 
high accuracy. However it has some limitations on memory usage which stem from 
choosing the size of the era. 

The TSBTS algorithm also provides some benefits for parallelization of $N$--body 
algorithms. Development of parallel versions of variable time--step codes becomes 
increasingly necessary for many areas of research, such as stellar dynamics in 
astrophysics, plasma simulations in physics, and molecular dynamics in chemistry 
and biology. The most natural way to do this is through the use of block time--steps, 
where each particle has to choose its own power of two, for the size of its time--step
\citep{Aarseth-2003}. Block time--steps allow efficient parallelization, given that 
large numbers of particles sharing the same block time--step can then be integrated 
in parallel.

In Section 2, we summarize the TSBTS algorithm time-symmetric block time--step algorithm. 
We provide definitions for the era concept, and for time-symmetrization of block time--steps.  
In Section 3, we present sample numerical tests for choosing the size of the era. 
We show how important is the effect of the era size on the energy errors, and the relationship 
between era size and iteration number. In Section 4, we offer a dynamic era size scheme for 
both better energy conservation and better memory usage. In Section 5, we present
a parallel algorithm for the TSBTS scheme with a hybrid force calculation procedure.
In Section 6, we discuss load balance and parallel performance tests of the algorithm.
Section 7 sums up the study.

\section{Era-Based Iterative Time Symmetrization for Block Time Steps}
\label{sec:1}

In the TSBTS algorithm, an iterative scheme is combined with an individual block 
time--step scheme to apply the algorithm to the $N$--body problem effectively. 
There are two important points in this algorithm: the era concept and the 
time-symmetrization procedure. The era is a time period in which we collect 
and store information for all positions and velocities of the particles for every 
step. At the end of each era, we synchronize all particles with time symmetric 
interpolation. This synchronization is repeated many times during the integration 
period, depending on the size of the era. 

Let us remember the TSBTS algorithm briefly:

We used a self-starting form of the leapfrog scheme; 
\begin{eqnarray}
{\bf r}_{new} &=& {\bf r}_{old} + {\bf v}_{old} {\rm \Delta} t
 + \frac{1}{2} {\bf a}_{old} {\rm \Delta} t^2,            \nonumber\\
 {\bf v}_{new} &=& {\bf v}_{old} + \frac{1}{2} ({\bf a}_{old} + {\bf a}_{new}){\rm \Delta} t,
\label{leapfrog}
\end{eqnarray}

with Taylor expansion for predicted velocities and positions;
\begin{eqnarray}
 {\bf r}_{new}^p &=& {\bf r}_{old} + {\bf v}_{old}{\rm \Delta} t + \frac{1}{2}{\bf a}_{old}{\rm \Delta} t^2,   \nonumber\\
 {\bf v}_{new}^p &=& {\bf v}_{old} + {\bf a}_{old}{\rm \Delta} t.
\end{eqnarray}

One of the easiest estimates for the time--step criterion is the {\em collisional time--step}. 
When two particles approach each other, or move away from each other, the ratio between 
relative distance and relative velocity gives us an estimation.

On the other hand, if particles move at roughly the same velocity, the collision 
time scale estimate produces infinity when the particles' relative velocities are zero. 
For such cases, we use a {\em free fall time scale} as an additional criterion, or 
just take the allowed largest time--steps for those particles.

Time-steps are determined using both the free-fall time scale and the collision time scale 
(\ref{eq:nbdt}) for particle $i$ by taking the minimum over the two criterion and over 
the all $j$ as;

\begin{equation}
\label{eq:nbdt}
\delta t_i = \eta \min_{i \neq j}\left\{ \frac{|r_{ij}|}{ |v_{ij}|},\sqrt{\frac{|r_{ij}|}{|a_{ij}|}}\right\}
\end{equation}

where $\eta$ is a constant accuracy parameter, $r_{ij}$ and $v_{ij}$ are
the relative position and velocity between particles $i$ and $j$, and $a_{ij}$ is the 
pairwise acceleration.

Even if Aarseth's time--step criterion \citep{Aarseth-2003} serves us better in avoiding 
such unexpected situations and gives us a better estimation, it needs higher order 
derivatives and it is expensive for a second order integration scheme.

Our time-symmetry criterion is defined in Eq.\ref{blockcondition}. This 
criterion gives us the smallest $n$ values that suit the condition 
${\rm \Delta} t_n \le \delta t^m_{i}$;

\begin{equation}
n = \min_{k \ge 1}
\left\{
 k \ \Big| \ \frac{1}{2^{k-1}} \le \frac{(\delta t^m_i + \delta t^{m+1}_i)}{2}
 \right\}                                           \label{blockcondition}
 \end{equation}

where $m$ is the iteration counter. Here, $m$ and $m+1$ refer to the
beginning and end of the time step.

In the case of block time--step schemes, a group of particles advances at
the same time. At each step of the integration, a group of particles is  
integrated with the smallest value of ${\rm \Delta} t_n$. Here, we refer 
to the group of particles as particle blocks. The first group of particles 
in an era is called the {\em first block}. 

In the first pass through an era, we perform standard forward integration 
with the standard block step scheme, without any intention to make the 
scheme time--symmetric. 
To compute the forces on the particles with the smallest value of 
${\rm \Delta} t_n$, we use second-order Taylor expansions for the predicted 
positions, while a first-order expansion suffices for the predicted velocity. 
Predicted positions, velocities, and accelerations for each particle for every 
time--step are stored during each era.

In the second pass, which is the first iteration, instead of Taylor expansions 
we use time-symmetric interpolations with stored data. This time, each time--step 
is calculated in a different way for symmetrization as in Algorithm \ref{algorithm}.
Here, $dt_m$ is the block time--step of the integrated particle group, and 
$ {\rm \Delta} t_n$ is the $n$'th level block time--step, which is obtained from 
a time-symmetry criterion (Eq.\ref{blockcondition}). If the current time is an 
even multiple of the current block time--step, that time value is referred to 
as {\em even time}, otherwise it is referred to as {\em odd time}.


\begin{algorithm}
\caption{ Symmetrization Scheme for Block Time Steps}
\label{algorithm}
\begin{algorithmic}
\FOR{$m = 1$ to number of iteration}
\IF{$time$ == \it{odd time}}
  \IF{$dt_m \ne {\rm \Delta} t_n$ }
\STATE $dt_m = dt_m / 2$ 
  \ENDIF		
\ENDIF		
\IF{$time$ == \it{even time}}
  \IF{$dt_m < {\rm \Delta} t_n$}
\STATE $dt_m = dt_m * 2$ 
  \ENDIF
  \IF{$dt_m == {\rm \Delta} t_n$}
\STATE $dt_m = {\rm \Delta} t_n$  
  \ENDIF
  \IF{$dt_m > {\rm \Delta} t_n$}
\STATE $dt_m = dt_m / 2$ 
  \ENDIF
\ENDIF		
\ENDFOR		
\end{algorithmic}
\end{algorithm}

Here is the description of the symmetrization scheme for block time--steps 
(as in Algorithm \ref{algorithm}):
\begin{itemize}
\item[]
If the current time is {\bf{\em odd}}, first, we try to continue with the 
same time--step.  If, upon iteration, that time--step qualifies according to the 
time-symmetry criterion (as in Eq.\ref{blockcondition}), then we continue to use 
the same step size that was used in the previous step of the iteration.  If not, we use a 
step size half as large as that of the previous time--step. 

\medskip

\item[]
If the current time is {\bf{\em even}}, our choices are: doubling the previous 
time--step size; keeping it the same; or halving it. We first try the largest 
value, given by doubling.  If Eq.\ref{blockcondition} shows us that this larger 
time--step is not too large, we accept it: otherwise, we consider keeping the 
time--step size the same.  If Eq.\ref{blockcondition} shows us that keeping the 
time--step size the same is okay, we accept that choice: otherwise, we simply 
halve the time--step, in which case no further testing is needed.
\end{itemize}

The same steps are repeated for higher iterations as in the first iteration.
The main steps of the integration cycle is given by Algorithm \ref{seq_algorithm}.


\begin{algorithm}
\caption{ Sequential Algorithm for TSBTS}
\label{seq_algorithm}
\begin{algorithmic}[1]
\STATE Initialization: \\
\hskip28pt	- Read initial position and velocity vectors from the source. \\
\hskip28pt	- Arrange size in the memory. \\
\hskip28pt	- Initialize particles' forces, time--steps, and next block times.\\
\hskip28pt	- Sort particles according to time blocks. \\
\STATE Start the iteration for the era.
\STATE Start the integration for the first block of the era.
\STATE Predict position and velocity vectors of all particles
for the current integration time. If this is the first step of the iteration, or 
if the time of the particle is smaller than the current time, do direct prediction: 
otherwise perform interpolation from the currently stored data.
\STATE Calculate forces on the active particles.
\STATE Correct position and velocity vectors of the particles in the block.
\STATE Update their new time--steps and next block time.\\ 
\hskip28pt	- After the first iteration, symmetrize new time steps according to Algorithm \ref{algorithm}.
\STATE Sort particles according to time blocks.
\STATE Repeat from Step 3 while current time is $\le$ time at the end of the era. 
\STATE Repeat from Step 2 until the number of the iteration reaches the iteration limit.
\STATE Repeat from Step 2 for the next era, until the final time is reached.
\STATE Write the outputs and finish the program.
\end{algorithmic}
\end{algorithm}

\section{Numerical Tests for the Size of the Era}
\label{sec:2}
The size of an era can be chosen as any integer multiple of the maximum allowed 
time--step. There is not any important computational difference between dividing the 
integration to the small era parts and taking the whole simulation in one big era. 
However some symmetrization routines such as adjusting the time--steps and interpolating 
the old data increase the computation time. Additionally, keeping the whole 
history of the simulation requires a huge amount of memory.

It is important to decide what is the most convenient choice for an era. We need 
to store sufficient information from the previous steps to adjust the time--steps 
with iterations. To avoid doing additional work and storing a uselessly large history, 
choosing  a large size for the era is not recommended. On the other hand, the era 
size must be large enough to store rapid and sharp time--step changes.

We made several tests with different Plummer model initial conditions, using different 
sizes of era. Units were chosen as standard $N$--body units \citep{Heggie-2003}, as the 
gravitational constant $G=1$, the total mass $M=1$ and the total energy is $E_{tot}=-1/4$. 
We limited the maximum time--step to $1/64$. The $\eta$ parameter was kept larger 
than usual to see the error growth in smaller time periods. The $\eta$ parameter 
was set as 0.1 for 100-body problems, and 0.5 for 500-body problems. The Plummer type 
softening length $\epsilon$ was taken as 0.01. Each system was integrated for every 
era size ($1,0.5, 0.25, 0.125, 0.0625 , 0.03125, 0.015625$) for 1000 time units.

\begin{figure}[h]
\begin{center}
  \FigureFile(120mm,120mm){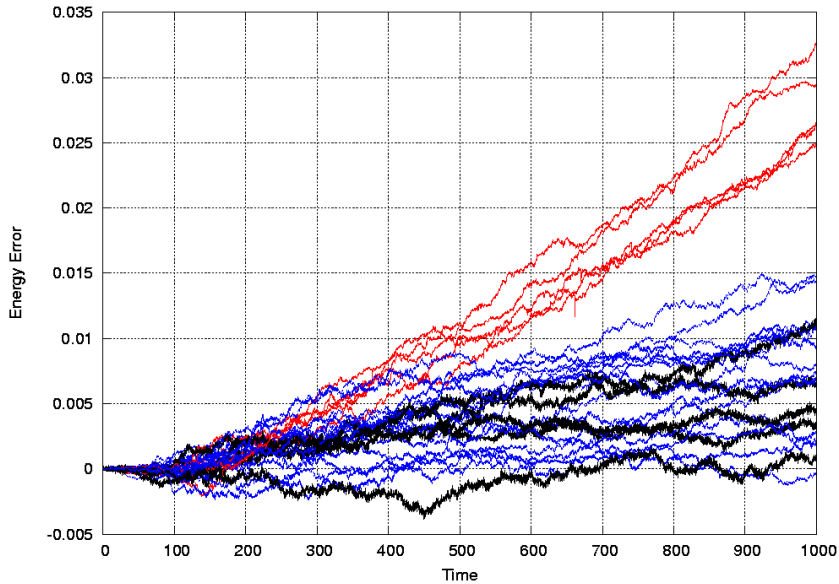}
\end{center}
\caption{Relative energy errors for 100-body problems. 5 different sets of 
Plummer model initial conditions with 5 different era sizes (0.015625, 0.03125, 
0.0625,0.125,0.25) are used with 3 iterations for 1000 time units. The top 5 
curves (red curves) show linearly growing errors that correspond to errors for 
the largest era sizes ($0.25$). The rest of the curves present the results for 
other era sizes. The smallest relative errors in the figure (black curves)
show a random-walk fashion and correspond to results to the smallest era size ($0.015625$).}
\label{fig1}
\end{figure}

Fig.\ref{fig1} shows the energy errors for 5 different 100-body problems with 
5 different era sizes. In these test runs, time-symmetrized block time--steps 
were used with 3 iterations. We also performed test runs for other era sizes 
($1.0, 0.5$). However, the growth of energy errors for these era sizes reached 
beyond the scales of this figure. The figure shows that, 3 iterations are not 
enough to avoid linearly growing errors for large (here, $0.25$) era sizes. 

\begin{figure}[h]
\begin{center}
\FigureFile(120mm,120mm){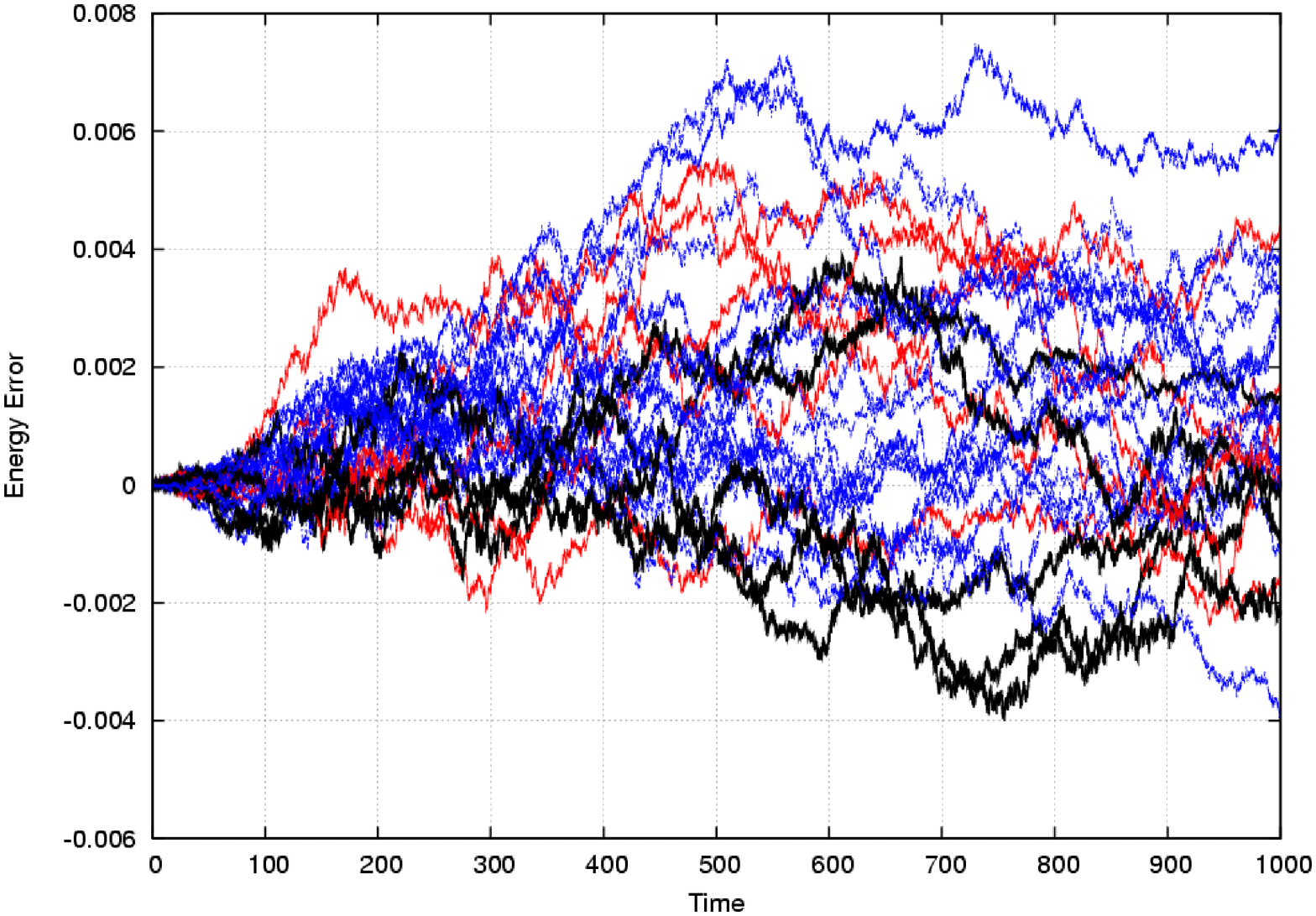}
\end{center}
\caption{Relative energy errors for 100-body problems. 5 different sets of 
Plummer model initial conditions are used for 5 iterations with 5 different 
era sizes (0.015625, 0.03125, 0.0625,0.125,0.25). In this figure, all of the 
curves show random-walk fashion instead of linearly growing errors. Also, the worst 
relative error is below $0.008$ even when it is $0.035$ in Fig.\ref{fig1}.}
\label{fig2}
\end{figure}

We conducted the  following tests to see this effect clearly. 
Fig.\ref{fig2} shows the energy errors for 5 different 100-body problems with 5 
different era sizes as in the previous figure. However, we used 5 iterations here. 
In this figure, the largest era size ($0.25$ time unit) does not show a linearly 
growing error exactly the contrary to the case of 3 iterations.The improvement on 
energy errors comes directly from the iteration process as we expected. 

\begin{figure}[h]
 \centering
  \FigureFile(120mm,120mm){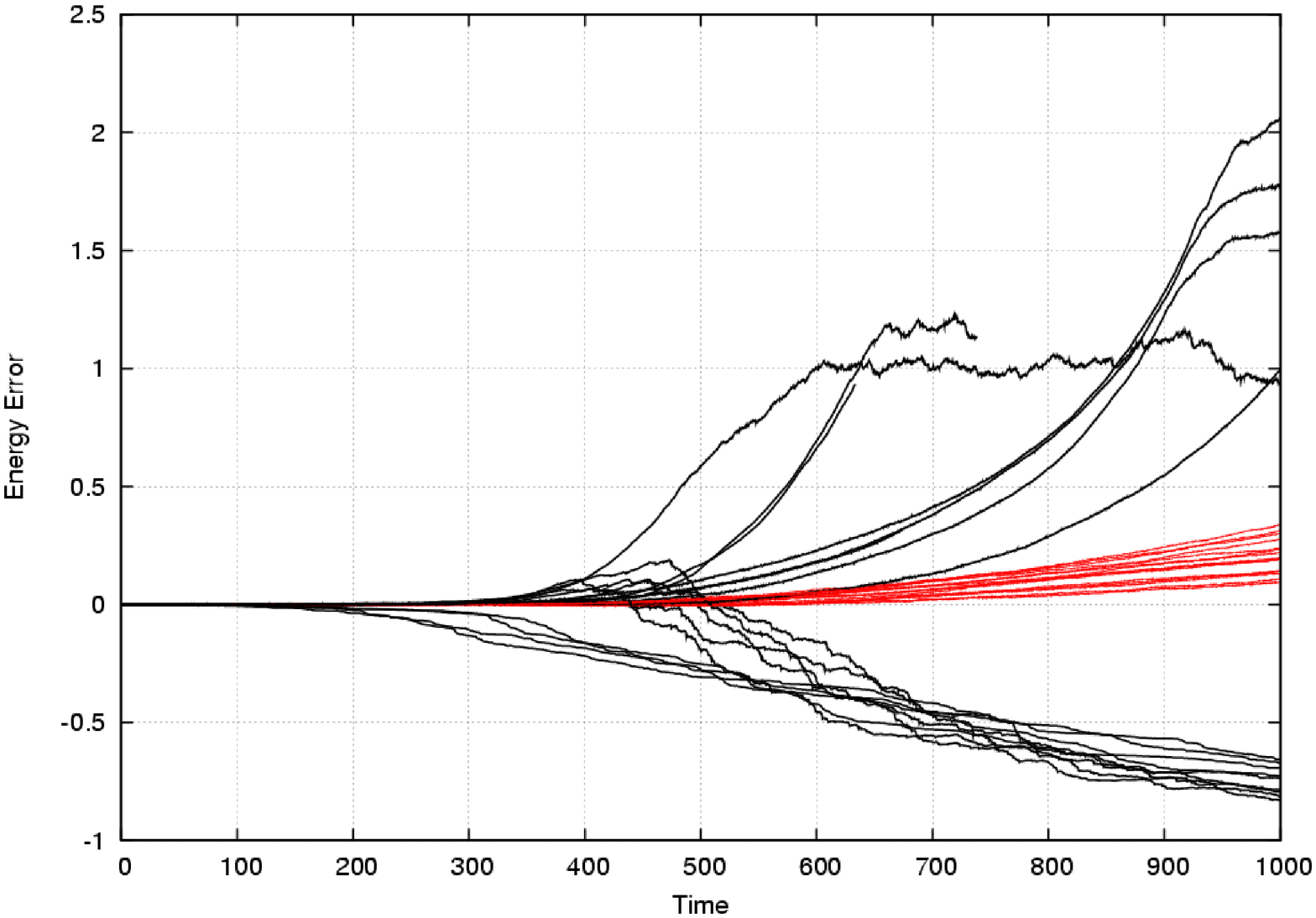}
   \caption{Relative energy errors for 500-body problems. 5 different sets of 
Plummer model initial conditions are used with 7 different era sizes (0.015625, 
0.03125, 0.0625, 0.125, 0.25, 0.5, 1) for each set. 3 iterations are performed 
in the integrations. 15 curves (red curves) in the center of the figure 
present the results of smaller era sizes ($ 0.015625, 0.03125, 0.0625$); the 
rest of the 20 curves correspond to larger ($ 0.125, 0.25, 0.5, 1$) era sizes.}
 \label{era_tests_p500_1}
\end{figure}

We increased the particle number 5 times, and set the $\eta$ parameter as $0.5$. 
The $\eta$ parameter could have been kept as $0.1$, but we forced the algorithm 
to take larger time--steps, which in turn produce  larger energy errors for 
relatively small time periods. Fig.\ref{era_tests_p500_1} shows the energy errors 
for 5 different 500-body problems with 7 different era sizes. The red curves show the 
errors for era sizes of $0.015625, 0.03125$, and $0.0625$ time units; the black 
curves show the errors for era sizes of $0.125, 0.25, 0.5$, and $1$ time units.

It seems that more iterations are needed to obtain  smaller energy errors while 
working with larger era sizes. If time-symmetric block time--steps can not be 
produced with a small number of iterations, the total energy error grows linearly. 
As indicated by our tests, iteration 
number and era size must be chosen carefully to ensure symmetric block time--steps.

Although the size of the era is not very important as long as the iteration number 
is large enough, a high number of iterations is not the preferred choice, as it 
demands high computational cost. Also, the era size would have to be kept small 
to avoid the huge memory usage. In practice, our tests show that, 5 iterations is 
not enough to prevent linearly growing errors when we use greater than $0.25$ time unit as the era size.

On the other hand, the era size must be greater than the greatest time--step. 
Otherwise we can not store past information for the iteration process and the algorithm 
works as a classical block time--step scheme. 

\section{Dynamic Era}
\label{sec:3}
Our test results for symmetrized time--steps with a small number of iterations in the 
previous section show that keeping the era size large or small has a clear effect on 
energy errors. However, the amount of the past position and the velocity information 
increase with the size of the era. Then, many more iterations are required to obtain 
optimized time--steps. And increased numbers of iterations consume more CPU time.

Let us remember and give some additional details and definitions about the relationship 
between block time--steps and era: similar to the {\em first block} definition we 
provided in Section 2, the last group of particles in an era is referred to as the 
{\em last block}. The current time in the integration for  the first and last blocks 
are referred to as {\em first block time} and {\em last block time}, respectively.

At the end of each era, integration of every particle stops at the same time, and 
new block time--steps are calculated and assigned for new blocks.  The last block can 
take the maximum allowed time--step at the most. The first block can take any block 
time--step smaller than the maximum allowed time--step. Then, particles are sorted 
according to their block time--steps. Also, every block has its own integration time 
related to its block time--step.

If we can find the proper criterion to change it, era size can be controlled dynamically.  
The simplest choices can vary between 1 time unit and the allowed largest time--step. 
Our suggestion is: calculate the new block time--steps and the first and last block 
times at the end of each era, and take 
the difference between the last and first block times. This difference gives us a 
dynamically changing size  and we can assign this as the size of the new era. 

Naturally, sometimes this difference can be larger than 1 time unit, or smaller 
than the maximum allowed time--step. Also, if all of the particles take the same time--step 
in any era, the difference goes to zero. We can use the maximum allowed time--step 
and any power-of-two times of this era size for the top and bottom limits of the era, 
respectively. 
Here, we used $2^{-3}$ multiples of the largest time--step for the lower limit. If 
all of the particles take the largest time--step, or larger time--steps than the new era 
size, there will not be enough past information for symmetrization. For these 
reasons, era size must not be much smaller than the largest time--step.

\begin{algorithm}
\caption{ Sequential Algorithm for TSBTS with Dynamic Era Size}
\label{seq_dynera_algorithm}
\begin{algorithmic}[1]
\STATE Initialization (same as Algorithm \ref{seq_algorithm}).\\
\STATE Set first and last block times.
\STATE Calculate dynamic era size ({\em dynamic era size} = {\em last block time} - {\em first block time}\\
 \hskip28pt	{\it i}) if {\em dynamic era size} $ < 2*10^{-3} *$ {\em maximum time step}\\
 \hskip56pt			{\em dynamic era size} $ = 2*10^{-3} *$ {\em maximum time step}\\
 \hskip28pt	{\it ii}) if {\em dynamic era size} $>$ {\em maximum time step}\\
 \hskip56pt			{\em dynamic era size} = {\em maximum time step}\\

\STATE Start the iteration for the era.
\STATE Start the integration for the first block of the era.
\STATE Predict position and velocity vectors of all particles
for the current integration time. If this is the first step of the iteration, or 
if the time of the particle is smaller than the current time, do direct prediction: 
otherwise perform interpolation from the currently stored data.
\STATE Calculate forces on the active particles.
\STATE Correct position and velocity vectors of the particles in the block.
\STATE Update their new time--steps and next block time.\\ 
\hskip28pt	- After the first iteration, symmetrize new time steps according to Algorithm \ref{algorithm}.
\STATE Sort particles according to time blocks.
\STATE Repeat from Step 5 while current time is $\le$ time at the end of the era. 
\STATE Repeat from Step 4 until the number of the iteration reaches the iteration limit.
\STATE Repeat from Step 2 for the next era, until the final time is reached.
\STATE Write the outputs and finish the program.
\end{algorithmic}
\end{algorithm}

If our estimate of the era size is smaller than our largest time--step, the particles with largest 
time--steps are excluded from the integration process of the era, and are then left 
for the next era. Errors of energy conservation oscillate in time, when they happen. 
We can use the allowed largest time--step for the era size in these cases.
The main steps of the algorithm is given by Algorithm \ref{seq_dynera_algorithm}.
 
\begin{figure}[h]
\centering
\FigureFile(120mm,120mm){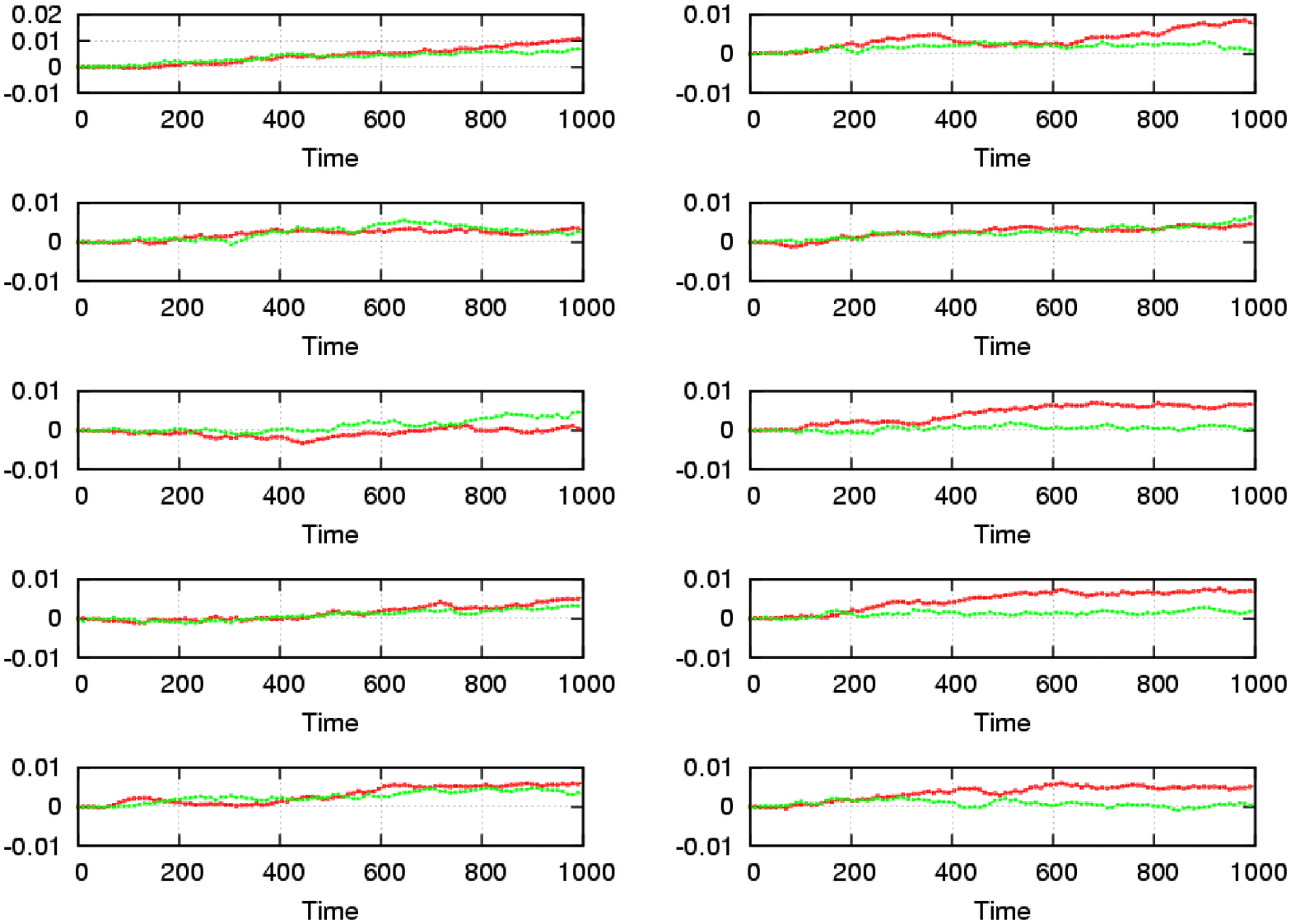}
\caption{Relative energy errors for 10 different 100-body problems.
For each initial condition, two algorithms are performed, one with fixed
and one with changing era size. 3 iterations are used for two algorithms.
Fixed era size was taken as $0.015625$. This value was also used as the allowed
largest time--step for the algorithms. The green curves correspond to dynamic 
era sizes and $70\%$ of them show smaller errors than fixed sizes.}
\label{fig:100dyn1}
\end{figure}

\begin{figure}[h]
\centering
\FigureFile(120mm,120mm){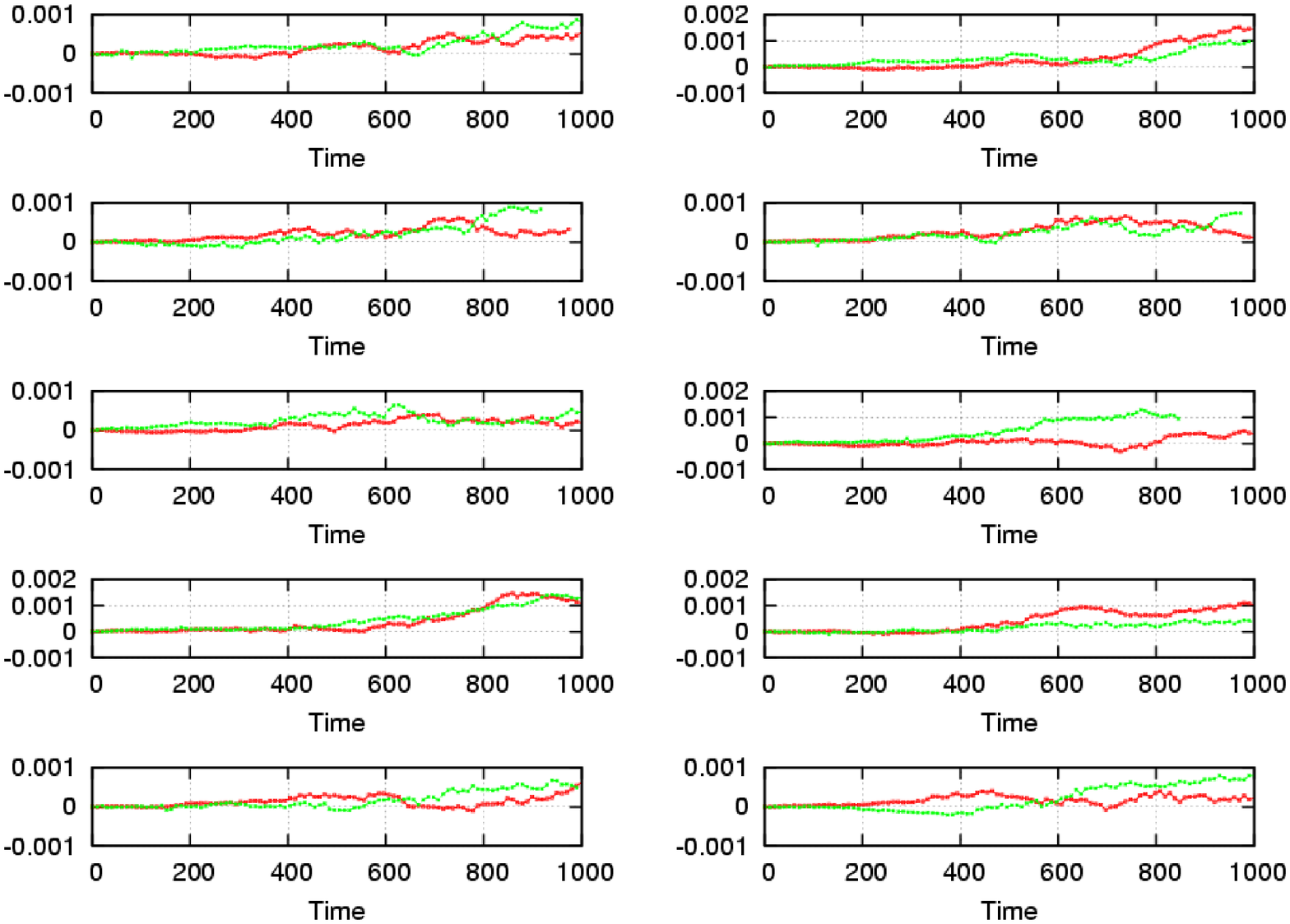}
\caption{Relative energy errors for 10 different 500-body problems. Fixed 
and dynamic era sizes are performed for each initial condition, as in 
Fig.\ref{fig:100dyn1}. Fixed era size and allowed largest time--step were taken 
as $0.015625$ just as in previous tests. The results for dynamic and fixed era 
sizes are in the same error ranges ($40\%$ of them show smaller errors than fixed 
sizes), and no linearly growing error is observed.}
\label{fig:500dyn1}
\end{figure}

In the tests we did for the dynamic era,  we used two choices for era size: equal to the allowed largest 
time--step, and dynamically changing size as defined above. We already know from 
previous runs for these test problems that we obtained the smallest errors on total 
energies when we took the allowed largest time--steps as the era size.
We performed 3 iterations. Fig.\ref{fig:500dyn1} shows the energy errors for 
10 different 500-body problems.  The green curves show the results for the dynamically 
changing era; the red curves show the results for the fixed era. Fig.\ref{fig:100dyn1} 
shows the energy errors for 10 different 100-body problems.

The results for dynamic era size are in the same range with those of fixed era size. 
Even if the chosen fixed era size ($0.015625$) seems like the best choice for previous 
tests with the same initial conditions and parameters (i.e., maximum allowed time--steps, 
softening and accuracy parameters), in general, dynamic era gives modestly better 
results than fixed era for $0.015625$. We ran more than 20 tests, and 
in $45\%$ of them were the errors for dynamic era size larger than errors for 
fixed era size. 
The rest of the results are clearly better than those for fixed era sizes, besides 
the advantage of reduced memory usage for the same number of iterations. 
Running times for dynamic era size are $10\% $ less than for fixed era sizes in general. 

\section{Parallel Algorithm }
\label{sec:4}
Basically, there are two well known schemes that are used in direct $N$--body 
parallelizations: copy and ring. 

The ring algorithm is generally preferred for reducing memory usage.
It can be reasonable for shared time--step codes, but it is not easy to 
use with block step schemes. It is also well known from previous works that 
this algorithm achieves almost the same speedup as the copy algorithm \citep{Makino-2002}.
The number of the particles in the integrated block changes with every step. 
In many cases, the size of the integrated block can be smaller than the number 
of the processors. It is difficult to obtain balanced load distribution for such cases. 

We used the copy algorithm. While it is much easier to extend for block step schemes,
the copy algorithm also has the load imbalance problem in classical usage. For any 
case, block size can be smaller than the number of processors again. 

We divided the partitioning strategy into two cases to avoid bad load balancing. 
In the first case, we divided the particles when the number of particles in the first 
block is greater than number of nodes. This is a kind of data partitioning, with 
every node containing a full copy of the system. In the second case, we divide the 
force calculation of the particles in the first block as a kind of work partitioning.

Our parallel algorithm works with the following steps, as in Algorithm \ref{par_algorithm}.


\begin{algorithm}
\caption{Parallel TSBTS Algorithm}
\label{par_algorithm}
\begin{algorithmic}[1]
 \STATE Broadcast all particles. Each node has a full copy of the system. 
 \STATE Initialize the system for all particles in all nodes. Every node computes 
 time--steps for all particles.
 \STATE Compute and sort time blocks. 
 \STATE Integrate particles in the first block whose block times are the minimum 
 for the era:\\
\hskip28pt	{\it i}) if the number of the first block $\ge$ number of nodes: every processor\\ 
\hskip28pt  calculates forces and integrates \\
\hskip28pt (number of first time block)/(number of nodes) particles.\\
\hskip28pt	{\it ii}) if the number of the first block $<$ number of nodes: every processor \\
\hskip28pt calculates (number of particles)/(number of nodes) part of the forces \\
\hskip28pt on the particles of the first block.\\
\STATE Update integrated particles.
\STATE Repeat from Step 3.
\end{algorithmic}
\end{algorithm}

\section{Load Balance and Performance}
\label{sec:5}
We have performed test runs on a Linux cluster in ITU-HPC Lab.\footnote{\.Istanbul Technical University High Performance Computing Lab.} 
with 37 dual core 3.40 GHz Intel(R) Xeon(TM) CPU with Myrinet interconnect.

The compute time was measured using MPI\_Wtime(). The timing for total compute time 
was started  before the broadcast of the system to the nodes, and ended at the end 
of integration. The calculation time of the subset of the particles in the current time 
block that are being handled by a given processor was taken as the work load of the
processor. 
In the iteration process, the largest time was taken as the work load of the processor 
for the same time block.

Work load of the $i$'th processor for every active integrated particle group is defined 
as $w_i$; $np$ is the number of processors; the mean work load $\left<W\right>$ is:

\begin{equation}
 \left<W\right> = \frac{1}{np}\sum_{i=1}^{np}w_i,
\end{equation}
and load imbalances:
\begin{equation}
 L(w)=1- \frac{\left<W\right>}{max(w_i)}.
\label{eq:imbalance}
\end{equation}

\begin{figure}[h]
\begin{center}
\FigureFile(120mm,120mm){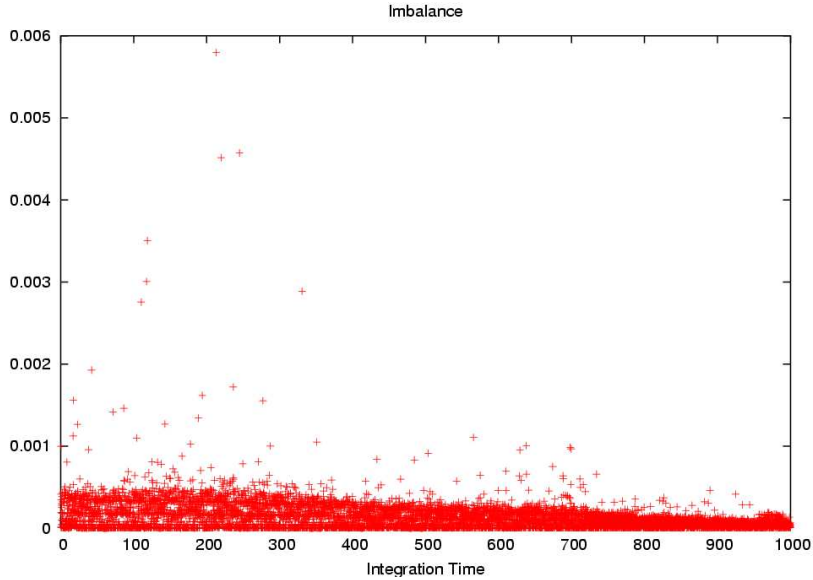}
\end{center}
\caption{Load imbalance for 1000-body problem  Plummer model initial conditions 
using 12 processors for 1000 time units. $\eta$ = 0.1; era size is taken as the 
allowed largest time--step. Every single red point corresponds to a load imbalance 
for the active particle group at the time when its vectors are updating.}
\label{fig:imbalance}
\end{figure}

Fig.\ref{fig:imbalance} shows the load imbalance for a 1000-body problem. 
We used 12 processors. In direct $N$--body simulations, a 1000 body is not 
a big number for 12 processors \citep{Makino-2002,Harfst-2007,Spinnato-2000}. 
Here, load imbalance is not seen as more than $0.1\%$ in general. Moreover,
load imbalance is smaller than expected. The main reason for this is in the 
iteration routines of the TSBTS algorithm, which increases both communication 
and calculation times for active particles. Also, when the number of particles in 
the first block is smaller than the number of nodes, work partitioning is applied 
in the algorithm, which also increases communication time.

$T_1$ is the running time for one processor; $T_n$ is the running time for $n$ processors.
${\it speedup}$, and ${\it efficiency}$ are given respectively, as:
\begin{equation}
 {\it speedup} = \frac{T_1}{T_n},
\end{equation}

\begin{equation}
 {\it efficiency} = \frac{T_1}{n * T_n}.
\end{equation}

\begin{figure}[h]
\begin{center}
\FigureFile(120mm,120mm){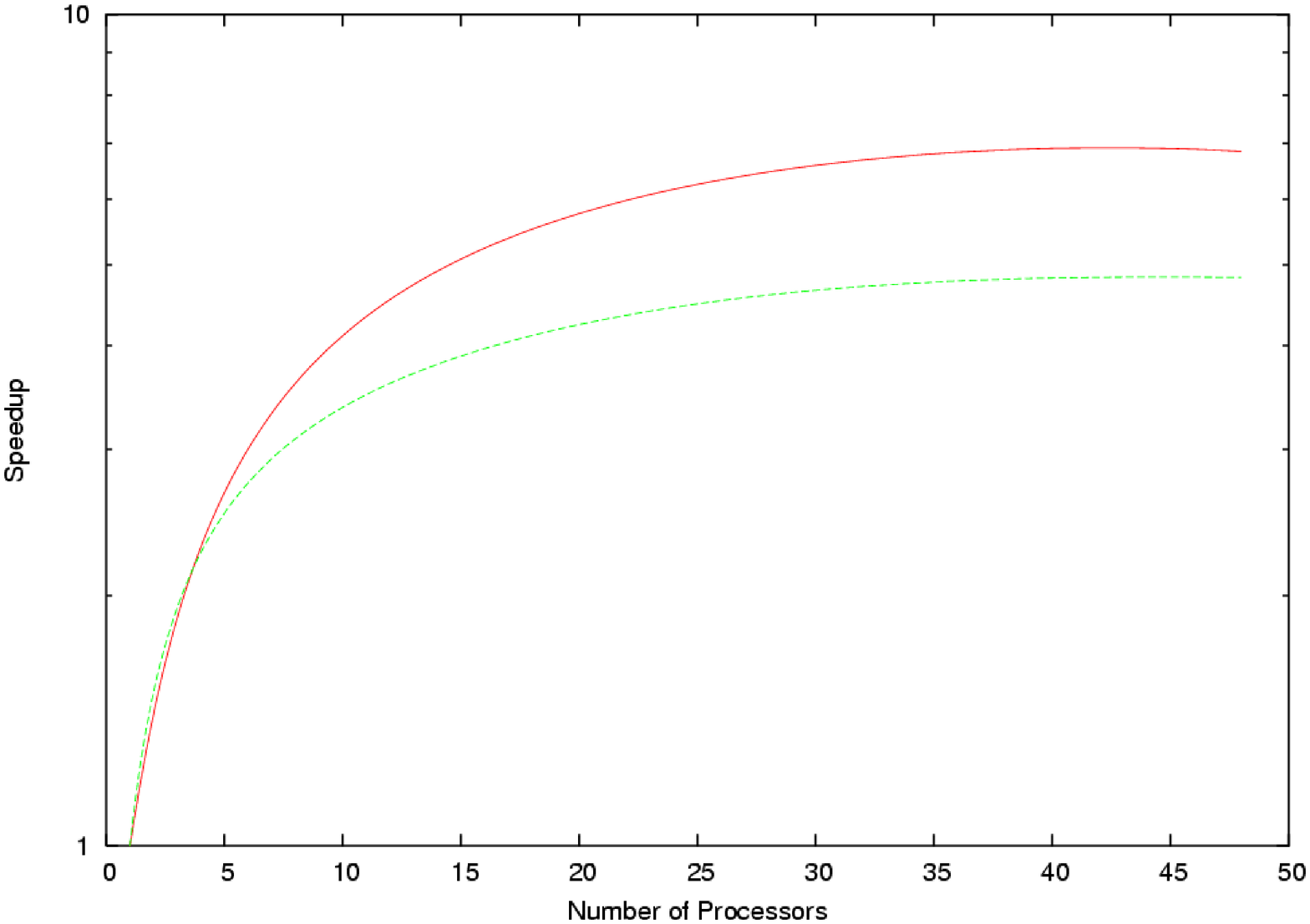}
\end{center}
\caption{Speedup vs. processor number for 10000-body problem Plummer model initial 
conditions, both for symmetrized and non-symmetrized individual block time--step 
algorithms. The continuous curve at the top corresponds to symmetrized block time--steps 
with 3 iterations. The discontinuous curve at the bottom corresponds to the classical 
block time--step algorithm.}
\label{fig:speedup}
\end{figure}

\begin{figure}
\begin{center}
\FigureFile(120mm,120mm){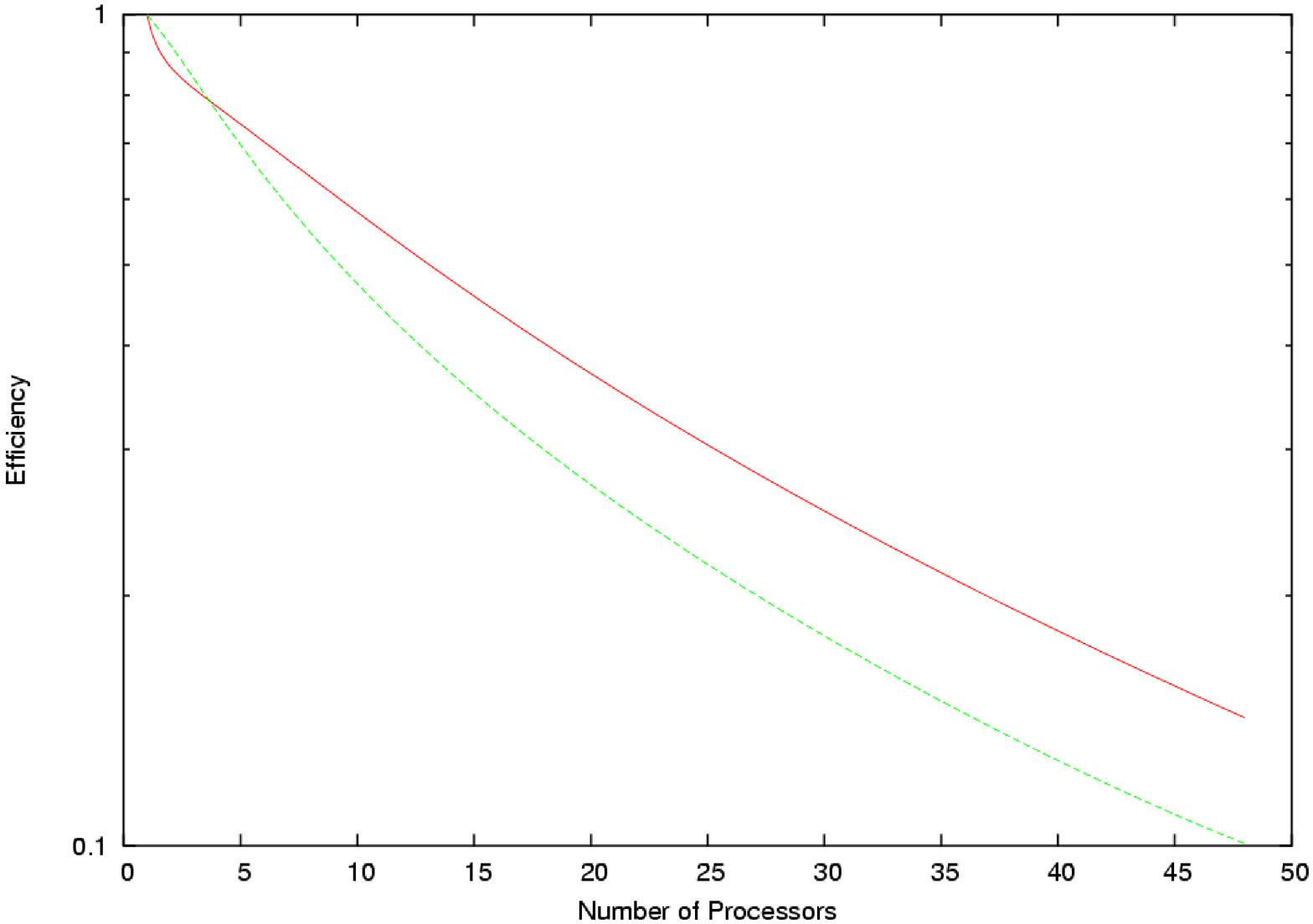}
\end{center}
\caption{Efficiency vs. processor number for 10000-body problem Plummer model initial 
conditions, both for symmetrized and non-symmetrized individual block time--step 
algorithms. The continuous curve at the top corresponds to symmetrized block time--steps 
with 3 iterations. The discontinuous curve at the bottom corresponds to the classical block
time--step algorithm.}
\label{fig:efficiency}
\end{figure}

Fig.\ref{fig:speedup} and Fig.\ref{fig:efficiency} show $speedup$ and $efficiency$ 
results of symmetrized and non-symmetrized block time--steps for an 10000-body problem 
initial conditions with Plummer softening length of $0.01$ and accuracy parameter 
$\eta=0.1$. Only one iteration with the TSBTS algorithm corresponds to individual 
block time--step algorithm without symmetrization. The speedup result for 3 
iterations is clearly better than the result for 1 iteration. These results
show that the communication/calculation ratio decreases with the iteration process,
though iteration needs much more computation time.

\begin{figure}[!h]
 \centering
   \subfloat[Energy Errors]{\label{fig:En-Err}\includegraphics[width=0.5\textwidth]{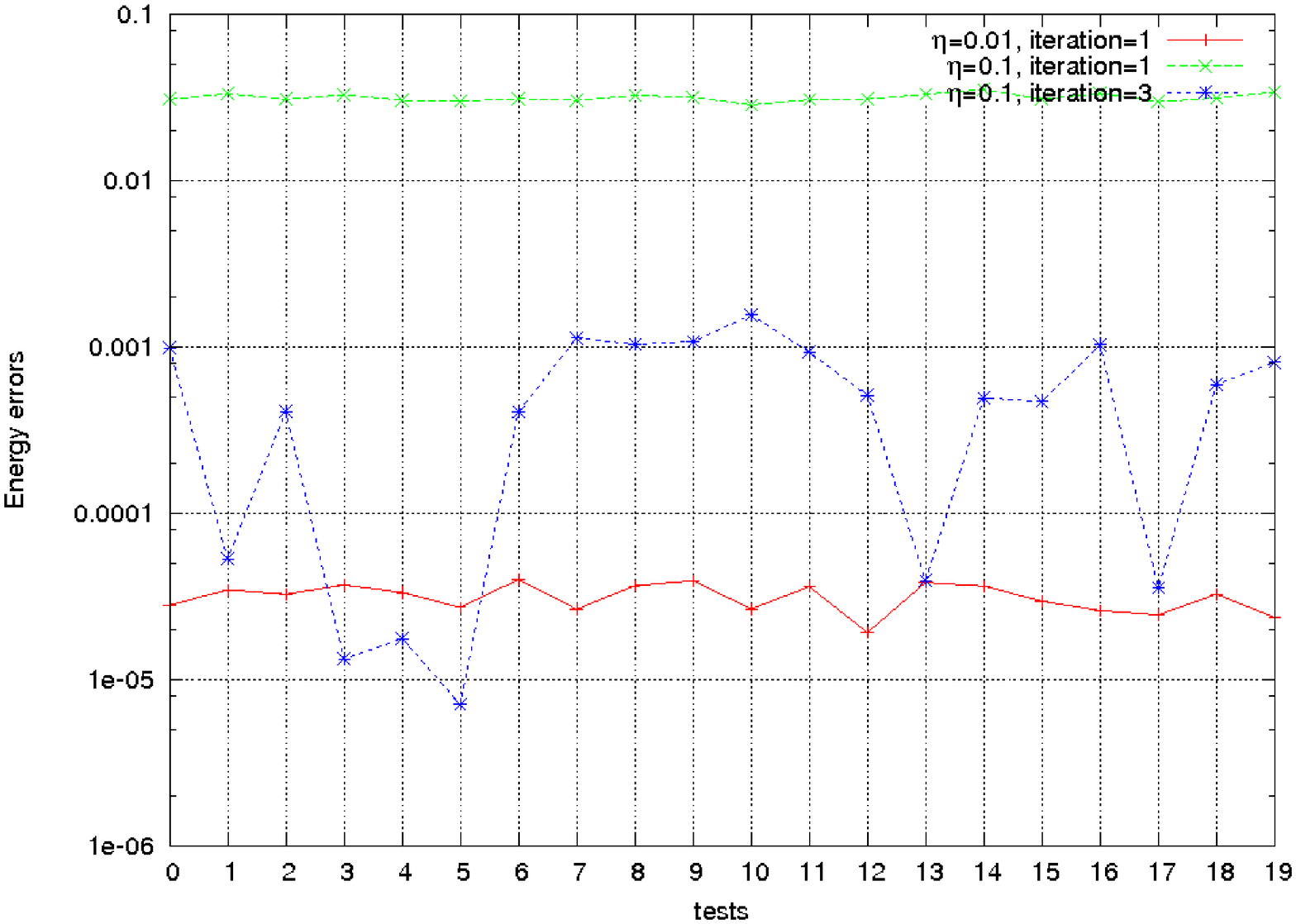}}
    \subfloat[CPU times]{\label{fig:CPU-times}\includegraphics[width=0.5\textwidth]{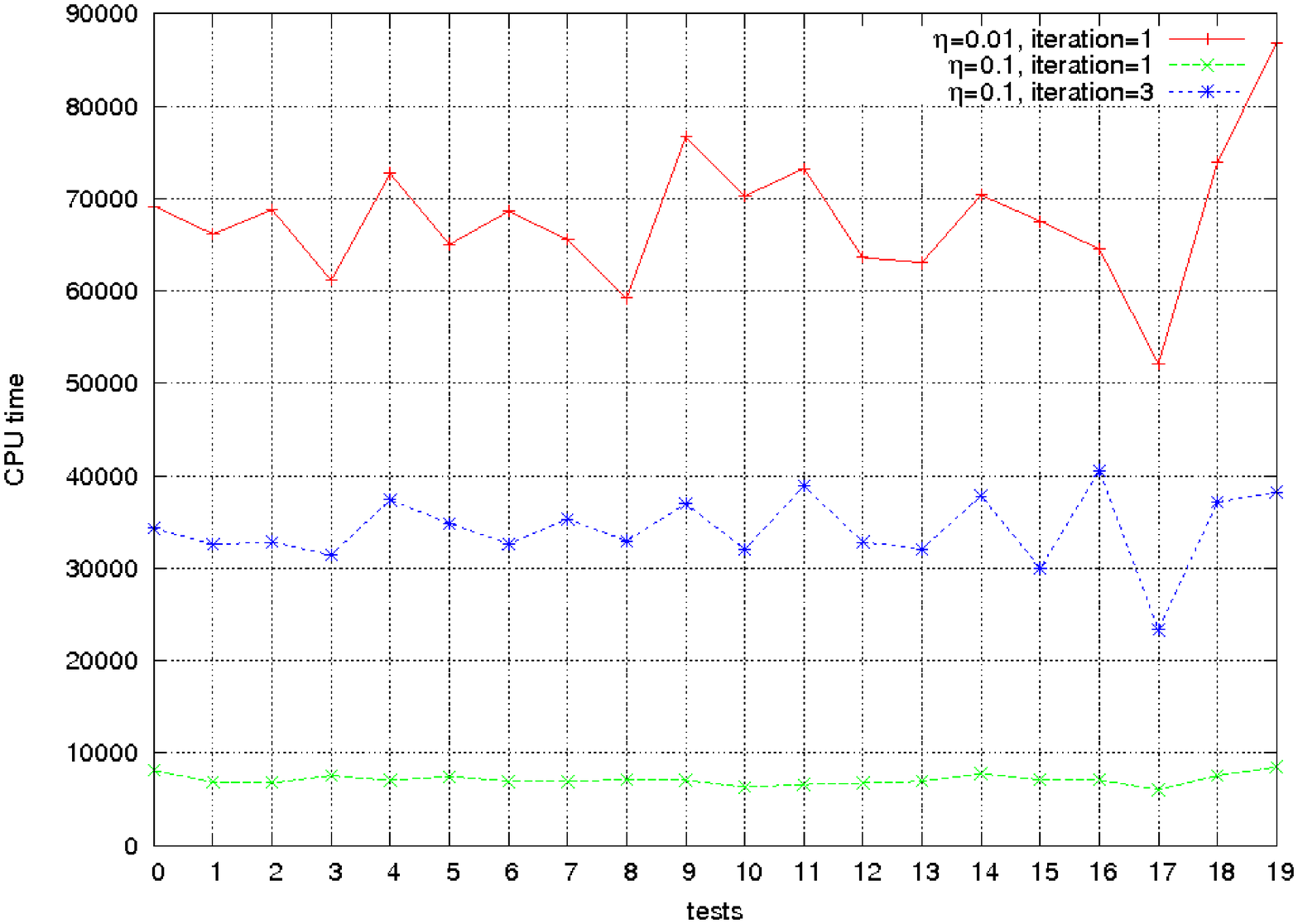}}
    \caption{Relative energy errors and CPU times for 20 different 500-body problems for 1000
    time units. One iteration with the TSBTS algorithm corresponds to the individual block 
    time--step algorithm without symmetrization. Here, we used two values for the accuracy 
    parameter ($\eta$ = 0.1, 0.01).}
 \label{fig:ErrCPU}
\end{figure}

For moderately short integration times, as in one time unit cases, the same error 
bounds can be obtained with less computation times by classical algorithms. 
However, the algorithm already shows its advantages in long time integrations. 
Fig.\ref{fig:ErrCPU} shows relative energy errors and CPU times for 20 different 
500-body problems with 2 different accuracy parameters ($\eta =(0.1,0.01)$) for 1 CPU. 
Each system was integrated for 1 and 3 iterations and 1000 time units. Even if it is 
not possible to obtain the same degree of energy errors for different test problems, 
the results are still highly promising. We obtained significantly better energy errors 
with the TSBTS algorithm (3 iterations) than with the classical individual block 
time--step algorithm (1 iteration) for the same accuracy parameters ($\eta = 0.1$) 
in all tests. Also, in some tests (more or less in $20\%$ of the tests), we obtained 
better results with 3 iterations for 10 times larger accuracy parameters than with 1 
iteration runs for $\eta = 0.01$.

For example in one of our 500-body problems, we obtained a relative energy error of 
$5.4*10^{-5}$ with $\eta = 0.1$ for 3 iterations, while it was $3.1*10^{-2}$ for 1 
iteration. To reach the same error bound with one iteration for 1000 time units, we 
had to reduce the accuracy parameter to 10 times smaller ($\eta=0.01$). 
Then, we obtained relative energy error of $1.92*10^{-5}$ with 1 iteration. In this 
example, calculation times for 1 and 3 iterations with $\eta = 0.1$  were 
$6.77*10^{3}$ sec., and $3.28*10^{4}$ sec. respectively, while the time was $6.36*10^{4}$ 
sec. for $\eta = 0.01$ with 1 iteration. Here, 3 iterations increase the calculation 
time by almost a multiple of 2. However, calculation time increases by a multiple of 
10, while the accuracy parameter is reduced by the same order.

\begin{figure}
 \begin{center}
  \FigureFile(120mm,120mm){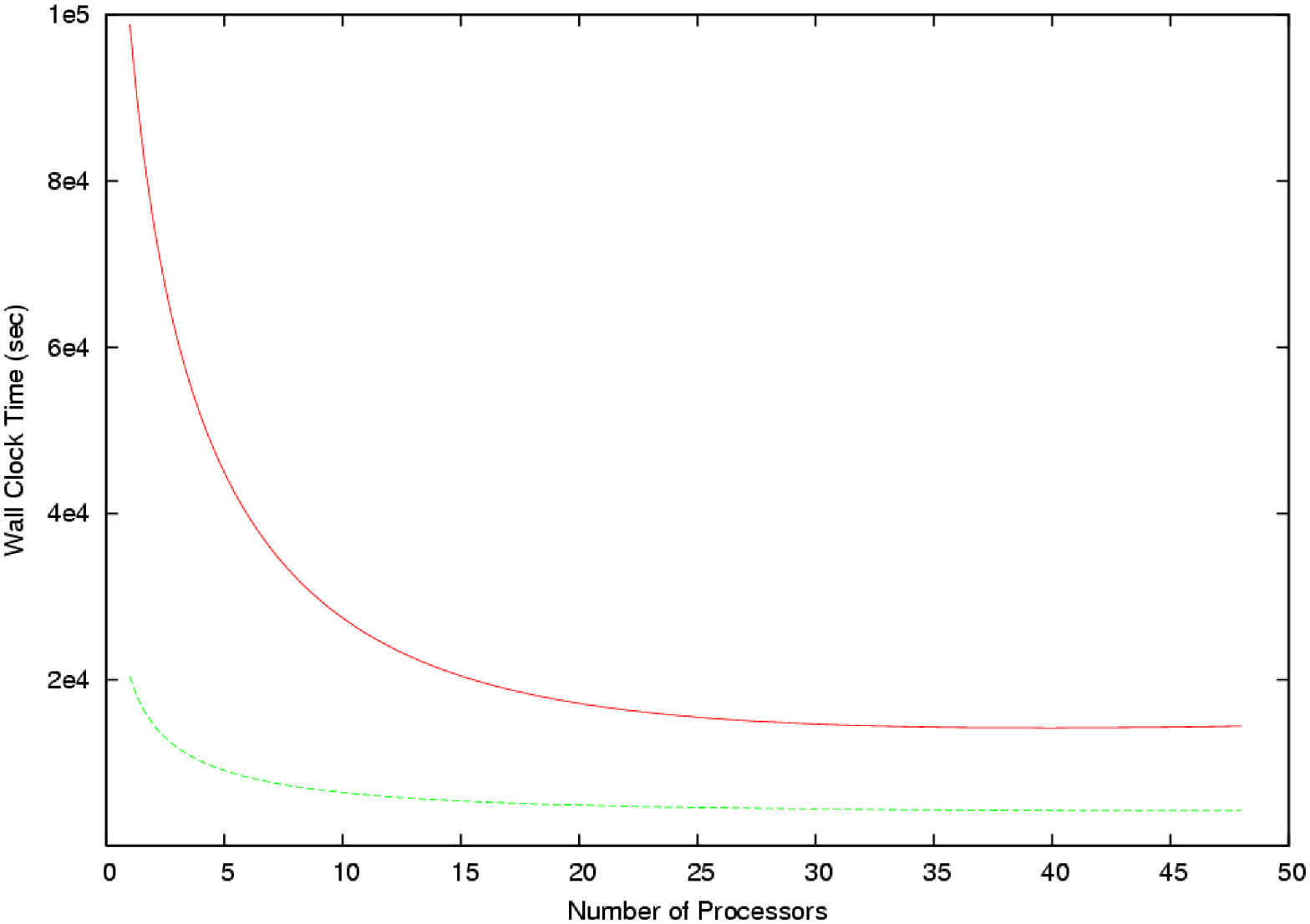}
 \end{center}
 \caption{Performance comparison of the TSBTS algorithm with 3 iterations 
and the classical individual block time--step algorithm for 10000-body problem 
Plummer model initial conditions. The top line corresponds to the TSBTS 
algorithm, and the one below corresponds to the non-symmetrized individual block 
time--step algorithm.}
 \label{fig:wallclocktime}
\end{figure}

Fig.\ref{fig:wallclocktime} shows running time requirements of the algorithm 
for the same 10000-body problem, both for 1 and 3 iterations, for one $N$--body 
time unit. The TSBTS algorithm needs up to 5 times more run time than 1 iteration case 
with 1 CPU for this test (for 500-body tests, this ratio was 4.75 as an average of their 
run times). 
This extra time is consumed by iteration and symmetrization procedures. 
The time-consuming ratio between the 1 and 3 iteration cases reduces to almost $3.5$ 
times when we increased the number of processors.

\section{Discussion}
\label{sec:6}
We have analyzed the era concept in greater detail for time symmetrized block 
time--steps. Our test results show that the size of the era must be chosen carefully. 
 This is important, especially for long-term simulations with highly desirable energy 
 conservations. The era size is also important to avoid the need for additional 
 data storage and a uselessly high number of iterations, which require too much 
 running time. 

In this work, we suggested a dynamically changing size for the era. This enables us
to follow the adaptively changing size for these time periods. In this scheme, the 
era size will be well-adjusted to the physics of the problem. In many cases, 
we obtained better energy errors than previous algorithm with fixed era size.

Additionally, we produced a copy algorithm-based parallel scheme combining with our 
time symmetrized block time--step scheme. We divided the force calculation into two 
approaches, according to the number of the integrating particles, to avoid bad load 
balancing. If the number of particles in the integrated block was greater than the 
number of processors, we used the classical approach --the copy algorithm-- to 
calculate forces. If we had a lower number of particles than processors to integrate, 
we divided the force calculations between the processors using work partitioning. 

Parallelization of direct $N$--body problem already features some difficulties 
regarding communication costs. Communication times dramatically increase with the 
number of processors. Previous works show that, using more than 10 processors for a 
few thousands particles does not result in a substantial gain \citep{Makino-2002,Harfst-2007,Spinnato-2000}.
This problem is replicated in individual time--step and block time--step cases. 

Even if we need to expend some additional communication efforts in our work partitioning 
approach, we obtain  good load balancing results with this approach. Also, the iteration 
process requires much more effort. 
Speedup and efficiency results are as we expected for current implementations. Scaling 
of the algorithm can be increased by using hyper systolic or other efficient algorithms 
\citep{Makino-2002} in future works. 

\bigskip

We thank the anonymous referees for their constructive comments which helped us to 
improve the contents of this paper. 
We acknowledge research support from ITU-HPC Lab. grant 5009-2003-03. 



\begin{thebibliography}{}

\bibitem[{Aarseth(2003)}]{Aarseth-2003}
Aarseth S J (2003) Gravitational $N$-Body Simulations. Cambridge University
  Press.

\bibitem[{Harfst et~al(2007)Harfst, Gualandris, Merritt, Spurzem, Zwart, and
  Berczik}]{Harfst-2007}
Harfst S, Gualandris A, Merritt D, Spurzem R, Zwart S, Berczik P (2007)
  Performance analysis of direct $N$--body algorithms on special-purpose
  supercomputers. New Astronomy 12:357--377

\bibitem[{Heggie and Hut(2003)}]{Heggie-2003}
Heggie D, Hut P (2003) The Gravitational Million-Body Problem. Cambridge
  University Press

\bibitem[{Makino(2002)}]{Makino-2002}
Makino J (2002) An efficient parallel algorithm for O($N^{2}$) direct summation
  method and its variations on distributed-memory parallel machines. New
  Astronomy 7:373--384

\bibitem[{Makino et~al(2006)Makino, Hut, Kaplan, and Sayg{\i}n}]{Makino-2006}
Makino J, Hut P, Kaplan M, Sayg{\i}n H (2006) A time-symmetric block time--step
  algorithm for $N$--body simulations. New Astronomy 12:124--133

\bibitem[{P.~Spinnato(2000)}]{Spinnato-2000}
Spinnato PS, van~Albada GD and Sloot PMA (2000) Performance analysis of parallel $N$--body
  codes. Proceedings of High Performance Computing and Networking, Lecture Notes 
in Computer Science v: 1823 p: 249-260 
\end{thebibliography}
\end{document}